%% file: CALL.TEX
\documentclass [winedt,yap]{iitparc}
\usepackage{cite}
\usepackage{amsmath,amssymb,amsfonts,bm}
\usepackage{eufrak}
\usepackage{lscape}
\usepackage{floatfig,wrapfig,epsfig}
\usepackage{subfigure}
\usepackage{color}
\usepackage{psboxit}
\usepackage{rotating}
\usepackage{curves}
\usepackage{ulem}
\usepackage[mathscr]{eucal}
\RequirePackage{psfrag}
\RequirePackage{graphicx}

\begin{document}\normalem
\initfloatingfigs
\frontmatter          

\IssuePrice{25.00}%
\TransYearOfIssue{2015}%
\TransCopyrightYear{2015}%
\OrigYearOfIssue{2015}%
\OrigCopyrightYear{2015}%

\TransVolumeNo{76}%
\TransIssueNo{8}%
\OrigIssueNo{8}%


\mainmatter

\setcounter{page}{161}
\CRubrika{NAVIGATION AND CONTROL OF MOVING SYSTEMS}
\Rubrika{NAVIGATION AND CONTROL OF MOVING SYSTEMS}

\include{AgaChe}

\end{document}

%% file: AgaChe.tex
\def\T{{\rm T}}
\def\R{\mathbb{R}}                               
\def\C{\mathbb{C}}                               
\def\ind{\mathop{\rm ind}\nolimits}              
\def\trace{\mathop{\rm trace}\nolimits}          
\def\NN{\mathop{\mathcal N}\nolimits}            
\def\RR{\mathop{\mathcal R}\nolimits}            
\def\TT{\mathop{\mathcal T}\nolimits}            
\def\sqa{\sqcap\!\!\!\!\sqcup}
\def\epr{\hfill$\sqa$}                 
\def\liml {\mathop{\lim}  \limits}               
\def\Pbes{P^\infty}                              
\def\G{\Gamma}                                   
\def\la{\lambda}                                 
\def\cdc{,\ldots,}                               
\def\1n{1,\ldots,n}                              
\def\LT {\tilde L}                               
\def\vj{\mathop{\tilde{J}}\nolimits}
\def\J {\mathop{\bar{J}}\nolimits}               
\def\xz{\hspace{-.07em}}
\def\xy{\hspace{.07em}}
\def\P{\widetilde P}                              

\def\spann{\operatorname{span}}                  

\title{The projection method for continuous-time consensus seeking\thanks{This work was supported by the Russian
Foundation for Basic Research, projects no.~13-07-00990, 13-01-13105, and 13-07-13167.}}
\titlerunning{The Projection Method for Continuous-time Consensus Seeking}

\author{R. P. Agaev and P. Yu. Chebotarev}
\authorrunning{Agaev, Chebotarev}
\OrigCopyrightedAuthors{R.P. Agaev and P.Yu. Chebotarev}
\institute{Trapeznikov Institute of Control Sciences, Russian Academy of Sciences, Moscow, Russia}

\received{Received October 22, 2014}

\maketitle

\begin{abstract}
For the case where the dependency digraph has no spanning in-tree, we characterize the region of convergence of the basic continuous-time distributed consensus algorithm and show that consensus can be achieved by employing the method of orthogonal projection, which has been proposed for the discrete-time coordination problem.
\end{abstract}

\section{INTRODUCTION}

For the coordination algorithms in networked multi-agent systems, conditions of reaching consensus are usually formulated in terms of spectral properties of the Laplacian matrix of the dependency digraph. In particular, it is well known that for the basic continuous-time protocol, asymptotic consensus is achieved for any initial conditions if and only if $0$ is a simple eigenvalue of the Laplacian matrix. A necessary and sufficient condition of this is \cite{AgaChe00} the presence of a spanning in-tree in the dependency digraph.
If this condition is satisfied, then consensus can be expressed \cite{CheAga09ARC,Olfati-SaberMurray04,MesbahiEgerstedt10book} by the inner product of the left eigenvector corresponding to the zero eigenvalue of the Laplacian matrix and the vector of initial opinions.
In~\cite{CheAga13UBSr,CheAga14IEEETAC}, it was found that for an arbitrary dependency digraph, the limiting state vector of the above protocol is equal to the product of the eigenprojection of the Laplacian matrix, $L,$ and the vector of initial opinions. The eigenprojection of $L$ coincides with the stochastic matrix of maximum in-forests of the weighted dependency digraph corresponding to the protocol (the forest consensus theorem). An analogous result for the discrete-time protocol by DeGroot involves Ces\`aro averaging.

In~\cite{AgaChe11ARCE1}, we proposed the \emph{orthogonal projection method\/} for reaching consensus in DeGroot's protocol with a proper (but not necessarily regular) stochastic dependency matrix. In~\cite{CheAga13UBSr}, it has been conjectured that this method can be applied to the basic continuous-time distributed consensus algorithm as well. The present paper verifies this conjecture.

In~Section\:\ref{s_notat090213}, we introduce the main notations and preliminaries. Other terminology used in this paper can be found in~\cite{AgaChe11ARCE1}.

\section{BASIC CONCEPTS AND PRELIMINARY RESULTS}
\label{s_notat090213}

\subsection{The continuous-time distributed consensus algorithm}
\label{s_model}

Consider the basic continuous-time algorithm of consensus seeking in a multi-agent system \cite{Olfati-SaberMurray04,CheAga09ARC}:
\begin{eqnarray}
\label{e_model1}
\dot x_i(t)
&\!=\!&u_i(t),\\
u_i(t)
&\!=\!&-\sum_{j=1}^na_{ij}\left({x_i(t)-x_j(t)}\right),\quad i=\1n,
\label{e_model2}
\end{eqnarray}
where $x_i(t)$ is the state (opinion) of agent $i$ and $a_{ij}\ge0$ is the weight with which agent $i$ takes into account the discrepancy with agent~$j.$ $A=(a_{ij})$ is the \emph{dependency matrix\/} of this algorithm.

The matrix form of the algorithm \eqref{e_model1}--\eqref{e_model2} is:
\begin{gather}
\label{e_modeL}
\dot x(t)=-L\,x(t), 
\end{gather}
where $x(t)=(x_1(t)\cdc x_n(t))^\T,$ $L$ is the \emph{Laplacian matrix of the algorithm\/} \eqref{e_model1}--\eqref{e_model2}
defined by
\begin{gather}\label{e_L}
L=\diag(A\xy\bm1)-A,
\end{gather}
and $\bm1=(1\cdc 1)^\T.$

${A=(a_{ij})}$ determines the weighted \emph{dependency digraph\/} $\G$ with vertex set ${V(\G)=\{\1n\}}:$ $\G$ contains arc $(i,j)$ with weight ${w_{ij}=a_{ij}}$ whenever ${a_{ij}>0}$ (i.e., when agent $i$ depends on agent~$j$ or, in other words, agent~$j$ influences agent~$\,i$). Thus, arcs in $\G$ are drawn from dependent agents to the agents that influence them; the weight $w_{ij}$ of arc $(i,j)$ is the degree of dependence of $i$ on~$j$.

\subsection{The Laplacian matrix and the matrix of maximum \mbox{in-forests}}
\label{s_forests}

By definition, the Laplacian matrix $L$ of the algorithm \eqref{e_model1}--\eqref{e_model2} has zero row sums, therefore, it is singular and $\bm{1}=(1\cdc 1)^\T$ belongs to the null space of~$L$.

According to Ger\v{s}gorin's theorem, the real parts of all nonzero eigenvalues of the matrix $(-L)$ which specifies the algorithm \eqref{e_modeL} are strictly negative.

Suppose that ${A\in\C^{n\times n}}$ is an arbitrary square matrix, $\RR({A})$ and $\NN({A})$ being the range and the null space of~$A,$ respectively. Let ${\nu=\ind A}$ be the {\em index\/} of $A,$ i.e., the smallest ${{k}\in\{0,1,\ldots\}}$ such that ${\rank A^{k+1}=\rank A^k}$ (${A^0\equiv I},$ where $I$ is the identity matrix of order~$n$).

{\em The eigenprojection\/} of matrix $A$ {\em corresponding to the eigenvalue}\footnote{Or simply the \emph{eigenprojection of\/}~$A$.}~$0$ is a projection (i.e., an idempotent matrix) $Z$ such that $\RR(Z)=\NN(A^{\nu})$ and ${\NN}(Z)=\RR(A^{\nu})$. In other words, $Z$ is the projection {\em onto $\NN(A^{\nu})$ along $\RR(A^{\nu}).$}

It was shown in~\cite{CheAga02ap} that the eigenprojection $\vj$ of $L$ 
coincides with the normalized matrix $\J=(j_{ks})$ of \emph{maximum in-forests\/} of the digraph $\G$ corresponding to~$L$ (i.e., of our dependency digraph). The entries of $\J$ are
\begin{equation*}
j_{ks}=\frac{f_{ks}}f,\quad k,s=\1n,
\end{equation*}
where $f$ is the total weight\footnote{The weight of a digraph (e.g., of an in-forest) is the product of the weights of all its arcs.} of all maximum in-forests of digraph $\G$, $f_{ks}$ being the total weight of those of them that have $k$ belonging to a tree with root (sink!)~$s.$

\section{Consensus and forests}
\label{s_DeGroot}

Suppose that the dependency digraph of a multi-agent system has no spanning in-tree or, equivalently \cite{AgaChe00}, the Laplacian matrix $L$ of the algorithm \eqref{e_model1}--\eqref{e_model2} has multiple zero eigenvalues. Then \cite[Theorem\:3.12]{MesbahiEgerstedt10book} there are vectors of initial opinions such that the algorithm \eqref{e_model1}--\eqref{e_model2} does not lead to consensus. 

\begin{theorem}[the forest consensus theorem \cite{CheAga13UBSr,CheAga14IEEETAC}]
\label{th_main}
Let $x(t)$ be a solution to the system~\eqref{e_modeL}. Then
\begin{gather}
\label{e_AsympState}
\lim_{t\to\infty}x(t)=\vj x(0), 
\end{gather} 
where $\vj$ is the the eigenprojection of~$L$ coinciding with the matrix of maximum in-forests $\J$ of the dependency digraph~$\G$.
\end{theorem}

Theorem\:\ref{th_main} can be derived from the expression ${x(t)=e^{-Lt}x(0)}$ for the solution of the system of equations \eqref{e_modeL} and the identity (see~\cite{CheAga14IEEETAC})
\begin{gather}
\label{e_liJ}
\vj=\lim_{t\to\infty}e^{-Lt}.
\end{gather}

The result of discretization of the algorithm \eqref{e_modeL} coincides (see~\cite{CheAga13UBSr}) with DeGroot's iterative pooling process \cite{DeGroot74}
\begin{gather}
\label{e_Pm}
y(k)=P^ky(0),\quad k=1,2,\ldots, 
\end{gather} 
where $y(k)$ is the state vector at discrete time $k,$ $P$ is the row stochastic matrix
\begin{gather}
\label{e_P}
P:=I-\tau L,
\end{gather} 
and $\tau>0$ is a sufficiently small parameter.

Observe that the criterion of stochasticity of the matrix \eqref{e_P} is
\begin{gather}
\label{e_tau}
0<\tau\le\left(\max_i\sum_{j\ne i}a_{ij}\right)^{-1}.
\end{gather} 

Let us compare the asymptotic properties of the algorithms \eqref{e_modeL} and~\eqref{e_Pm}. 
A necessary and sufficient condition of the convergence of $\{P^k\}$ is the aperiodicity of~$P$. On the other hand, the Ces\`aro limit
\begin{gather}
\label{e_Pinfry}
\Pbes:=\lim_{k\to\infty}\frac{1}{k}\sum\limits_{i=1}^k{P^i} 
\end{gather} 
exists for any stochastic matrix $P$ and coincides with $\,\lim_{k\to\infty}P^k$ whenever the the latter limit exists. Otherwise, if $P$ is periodic with period $s$, then $\Pbes=s^{-1}\!\left(P^{(1)}+\ldots+P^{(s)}\right),$ where $P^{(1)},\dots,P^{(s)}$ are the limits of the convergent subsequences of $\{P^k\}$: $P^{(i)}=\lim_{j\to \infty}P^{js+i}$, $\,i=1\cdc s.$

The following theorem is the discrete-time counterpart of Theorem~\ref{th_main}; it can be proved \cite{CheAga13UBSr} by combining a number of known results.
\begin{theorem}\label{th_main1} 
Suppose that the sequence $y(k)$ satisfies \eqref{e_Pm}\textup, where $P$ is determined by \eqref{e_P}--\eqref{e_tau}. Then
\begin{gather}
\label{e_AsympStateD1}
\lim_{k\to\infty}\frac{1}{k}\sum\limits_{i=1}^ky(i)=\vj y(0), 
\end{gather} 
where $\vj$ is the the eigenprojection of~$L$ coinciding with the matrix~$\J$ of maximum in-forests of the dependency digraph\/ $\G$ corresponding to~$L$. Furthermore\/\textup, if \eqref{e_tau} is satisfied strictly\/\textup, then
\begin{gather}
\label{e_AsympStateD2}
\lim_{k\to\infty}y(k)=\vj y(0).
\end{gather}
\end{theorem}

Thus, the main difference between Theorems~\ref{th_main} and~\ref{th_main1} is the necessity to use the Ces\`aro limit when the parameter $\tau$ of \eqref{e_P} takes its maximum value~\eqref{e_tau}.

\section{The consensus domain}
\label{s_DomaCons}

The consensus domain of a coordination algorithm is the set of initial vectors that are led (asymptotically) by this algorithm to vectors with equal components, i.e., to consensus vectors. The consensus domain of the algorithm \eqref{e_modeL} with matrix $L$ will be denoted by~$\TT(L)$.

\begin{theorem}\label{th_domain} 
For the algorithm \eqref{e_modeL} with matrix $L,$ it holds that $\TT(L)=\RR(L)\oplus\spann(\bm1),$ where $\spann(\bm1)$ is the linear span of the vector $\bm1=(1\cdc 1)^\T$.
\end{theorem} 

The proofs of Theorems\:\ref{th_domain}--\ref{th_ProFo} are given in the Appendix.

Thus, the basic continuous-time algorithm of consensus seeking \eqref{e_modeL} and the corresponding DeGroot's iterative pooling process \eqref{e_Pm} have the same consensus domain ${\TT(L)=\RR(L)\oplus\spann(\bm1)}$ (cf.\ \cite[Theorem~1]{AgaChe11ARCE1}),
except for the case where $P$ is not proper, which can only occur when ${\tau=\Bigl(\max_i\sum_{j\ne i}a_{ij}\Bigr)^{-1}},$ i.e., when $\tau$ reaches its upper bound~\eqref{e_tau}. In this case, only a generalized ``Ces\`aro consensus,'' as in~\eqref{e_AsympStateD1}, can be achieved. Under this generalization, the consensus domain is preserved, as the Ces\`aro limit coincides with the ordinary limit (Theorem\:\ref{th_main1}) whenever the latter exists.

\section{Orthogonal projection method for the continuous-time algorithm}
\label{s_Proj}

It follows from the theory of stochastic matrices that convergence to consensus in DeGroot's iterative pooling process \eqref{e_Pm} is guaranteed \emph{for any vector of initial opinions}~$y(0)$ 
if and only if~$P$ is regular\footnote{A stochastic matrix is called \emph{regular\/}~\cite{Gantmacher60} if it has no eigenvalues of modulus~1 except for the simple eigenvalue~$1.$}. For the more general case of a proper\footnote{A stochastic matrix is called \emph{proper}\/ if it has no eigenvalues of modulus $1$ except for $1.$} matrix $P,$ we have proposed \cite{AgaChe11ARCE1} the \emph{orthogonal projection method}, which leads to a \emph{quasi-consensus}. This method consists in: (1)~transformation of the vector of initial opinions into a vector belonging to the consensus domain of \eqref{e_Pm} $\,\TT'(L)=\RR(L)\oplus\spann(\bm1)$ by means of orthogonal projection and \,(2)~iterative correction of the latter vector by transformation~$P$.

Let us apply the method of orthogonal projection to the continuous-time consensus algorithm \eqref{e_model1}--\eqref{e_model2}. Then the system of equation \eqref{e_modeL} is preserved, while the vector of initial opinions is transformed, 
namely
\begin{gather}
\label{e_iniProj}
\tilde x(0)=Sx(0),
\end{gather}
where $S$ is the orthogonal projection onto the subspace $\TT(L)=\RR(L)\oplus\spann(\bm1)$. Then by Theorem\:\ref{th_main},
 \begin{gather}
\label{e_AsympStatee}
\lim_{t\to\infty}x(t)=\vj S\,x(0).
\end{gather}

Projection $S$ can be computed by means of the following proposition.
\begin{proposition}\label{p_STp} 
Orthogonal projection $S$ onto $\TT(L)=\RR(L)\oplus\spann(\bm1)$ has the representation
\begin{gather}
\label{e_projU}
S=U(U^\T U)^{-1}U^\T, 
\end{gather} 
where $U$ is any matrix obtained from $L$ by 
$(1)$~deleting one column corresponding to some vertex of each final class\/\footnote{A \emph{final class\/} is the vertex set of any bicomponent from which there is no arc directed outwards.} of the digraph $\G$ corresponding to $L$ and 
$(2)$~adding $\bm1$ as the first column.
\end{proposition}

\begin{example}
\label{ex_1}
Consider the multi-agent system with dependency matrix $A$ given below and find the corresponding matrices\, $L=\diag(A\xy\bm1)-A$\, and\, $U${\rm:} 

\begin{gather*}
A=\left(\begin{array}{ccccccc}
0 & 0 & 3 & 0 & 0 & 0 & 0\\
1 & 0 & 0 & 0 & 0 & 0 & 0\\
4 & 2 & 0 & 0 & 0 & 0 & 0\\
0 & 0 & 0 & 0 & 3 & 0 & 0\\
0 & 0 & 0 & 2 & 0 & 0 & 0\\
0 & 1 & 3 & 0 & 0 & 0 & 3\\
0 & 0 & 0 & 2 & 0 & 2 & 0\\
\end{array}\right),\quad
L=\left(\begin{array}{rrrrrrr}
 3& 0&-3& 0& 0& 0& 0\\
-1& 1& 0& 0& 0& 0& 0\\
-4&-2& 6& 0& 0& 0& 0\\
 0& 0& 0& 3&-3& 0& 0\\
 0& 0& 0&-2& 2& 0& 0\\
 0&-1&-3& 0& 0& 7&-3\\
 0& 0& 0&-2& 0&-2& 4\\
\end{array}\right),
\\[.6em]
U=\left(\begin{array}{rrrrrr}
 1& 0&-3&  0& 0& 0\\
 1& 1& 0&  0& 0& 0\\
 1&-2& 6&  0& 0& 0\\
 1& 0& 0& -3& 0& 0\\
 1& 0& 0&  2& 0& 0\\
 1&-1&-3&  0& 7&-3\\
 1& 0& 0&  0&-2& 4\\
\end{array}\right).
\end{gather*}

Using expressions $\vj=\lim_{\tau\to\infty}(I+\tau L)^{-1}$ $($see $(6)$ in {\rm \cite{AgaChe11ARCE1})} and \eqref{e_projU} one can obtain
\begin{gather*}
\vj\approx\left(\begin{array}{rrrrrrr}
0{.}4  &0{.}4  &0{.}2  &0      &0      &0&0\\
0{.}4  &0{.}4  &0{.}2  &0      &0      &0&0\\
0{.}4  &0{.}4  &0{.}2  &0      &0      &0&0\\
0      &0      &0      &0{.}4  &0{.}6  &0&0\\
0      &0      &0      &0{.}4  &0{.}6  &0&0\\
0{.}291&0{.}291&0{.}146&0{.}109&0{.}164&0&0\\
0{.}146&0{.}146&0{.}073&0{.}255&0{.}382&0&0\\
\end{array}\right),
\\[.6em]
S=
\frac1{22}\left(\begin{array}{rrrrrrr}
18&-4&-2& 4& 6& 0&0\\
-4&18&-2& 4& 6& 0&0\\
-2&-2&21& 2& 3& 0&0\\
 4& 4& 2&18&-6& 0&0\\
 6& 6& 3&-6&13& 0&0\\
 0& 0& 0& 0& 0&22&0\\
 0& 0& 0& 0& 0&0&22\\
\end{array}\right).
\end{gather*}
Since not all rows of $\vj$ are equal, consensus is not generally achieved. Let us compute
\begin{eqnarray}
\vj S &=& \bm1\!\cdot\!\frac1{110}\bigl(26      \;\;      26\;\;      13\;\;      18\;\;      27\;\;0\;\;0\bigr)\nonumber\\
 &\approx&\bm1\!\cdot\! \bigl(0{.}2364\;\;0{.}2364\;\;0{.}1182\;\;0{.}1636\;\;0{.}2455\;\;0\;\;0\bigr).
\label{e_Pinex}
\end{eqnarray}
The matrix $\vj S$ enables one to find the quasi-consensus $\vj S x(0)$ (see~\eqref{e_AsympStatee}) of the orthogonal projection method for the procedure \eqref{e_model1}--\eqref{e_model2}.
\end{example}

\section{An alternative form of the projection method}
\label{s_ProjA}

As was noted in Section\:\ref{s_Proj}, 
when the orthogonal projection method is applied to DeGroot's iterative pooling process, the vector of initial opinions $x(0)$ is corrected once by projecting onto $\TT'(L)$, i.e., it is replaced with the vector $S\,x(0)$, where $S$ is the projection \eqref{e_projU}; the subsequent correction is performed by the initial transformation~$P$. Therefore, the coordination algorithm as a whole is representable as
\begin{gather}
\label{e_DeGProj}
x(k)=P^kSx(0), \end{gather}
and the limiting state vector $x(\infty)$ in the case of a proper matrix $P$ has the form 
\[
x(\infty)=\Pbes Sx(0).
\]

Does there exist a matrix $\P$ such that the process \eqref{e_DeGProj} has an alternative representation
\begin{gather}
\label{e_DeGProjj}
x(k)=\P^{\xy k}x(0)\,? \end{gather}
In particular, is the matrix $PS$ suitable for the role of~$\P$? The following theorem gives positive answers to these questions.
\begin{theorem}
\label{th_AltDe} If
\begin{gather}
\label{e_Ptil}
\P=PS,
\end{gather}
then the coordination algorithm\:\eqref{e_DeGProjj} coincides with the orthogonal projection algorithm\/\:\eqref{e_DeGProj}.
\end{theorem} 

It can be observed that the row sums of $\P=PS$ are unity, since the projection $S$ does not alter the vector ${\bm1\in\TT(L)}$ (and thus, has row sums~$1$). On the other hand, $\P$ is not generally stochastic, since, as can be shown by examples, $\P$ may have negative entries.

Does there exist an analogous alternative form of the orthogonal projection method for the continuous-time procedure~\eqref{e_modeL}? To construct it, we first define $P$ by means of \eqref{e_P}--\eqref{e_tau}, then compute $\P=PS,$ and finally, having in mind \eqref{e_P}, find
\begin{gather}
\label{e_Ltil}
\LT:=\tau^{-1}(I-\P)=\tau^{-1}(I-(I-\tau L)S)=\tau^{-1}(I-S)+L S.
\end{gather}

Let us study the algorithm
\begin{gather}
\label{e_modeLL}
\dot x(t)=-\LT\,x(t)
\end{gather} 
which claims to be an alternative form of the projection method for the consensus protocol~\eqref{e_modeL}. This algorithm does not coincide with the projection method, since in the latter, projection \eqref{e_iniProj} onto $\TT(L)$ is performed ``abruptly,'' while the dynamics determined by \eqref{e_modeLL} is continuous. The matrix~$\LT$ is not generally Laplacian: having zero row sums (since $I$ and $\P$ have row sums~$1$), it, as can be shown by examples, may have strictly positive off-diagonal entries. Therefore, the terms of control $(-\LT\,x(t))$ (cf.\ \eqref{e_model1}) may not only bring agents' states closer, but also move them apart.

In the following theorem, we present some spectral properties of the matrix~$\LT$.
\begin{theorem}\label{th_SpeL} {\ }

{\rm 1.} Let $(\underbrace{0,\ldots,0}_d,\la_{d+1},\ldots,\la_n)$ be the spectrum of~$L.$
Then the spectrum of $\LT$ is $(0,\underbrace{\tau^{-1},\ldots,\tau^{-1}}_{d-1},$ $\la_{d+1},\ldots,\la_n).$

{\rm 2.} $\vj\xz S$ is the the eigenprojection of~$\LT.$

{\rm 3.} $\tau^{-1}$ is a singular value of $\LT$ with multiplicity \,$d-1.$
\end{theorem}

The following theorem is an analogue of the forest consensus theorem, as it gives an expression for the limiting state of the consensus algorithm~\eqref{e_modeLL}.
\begin{theorem}\label{th_ProFo} 
Let $x(t)$ be a solution of the system~\eqref{e_modeLL}. Then
\begin{gather}
\label{17042014eq1}
\lim_{t\to\infty}x(t)=\vj\xz S\xy x(0).
\end{gather}
\end{theorem}

By Theorem~\ref{th_ProFo}, the limiting states of the protocol~\eqref{e_modeLL}, can be expressed (as well as for the algorithm \eqref{e_modeL}) by the product of the eigenprojection of the matrix that determines the algorithm and the vector of initial opinions. Theorem~\ref{th_ProFo} is not a corollary of the forest consensus theorem, as $\LT$ is not generally a Laplacian matrix.

\begin{remark}
The identity $\lim_{t\to\infty}\exp({-\LT t})=\vj\xz S$ which implies Theorem~\ref{th_ProFo} can also be derived in a different way. Using the commutativity of $S$ and $LS$ (since $SL=L$ by Theorem\:\ref{th_domain}) and the definition~\eqref{e_Ltil} we have (provided that the limit in the right-hand side exists):
\begin{gather}
\label{eq_lim1}
  \lim_{t\to\infty}\exp\,(-\LT t)
 =\lim_{t\to\infty}\exp\left(-(I-S)\tau^{-1}t\right)\exp(-LSt).
\end{gather}

Using the equalities ${\exp(A)=\sum_{k=0}^\infty\frac{A^k}{k!}}$, ${S^k=S},$ and ${(I-S)^k=I-S},$
(${k=2,3,\ldots}$) it is straightforward to verify the identities
\begin{eqnarray*} 
\exp\left({-(I-S)t/\tau}\right)
&=&S+(I-S)\exp\left(-t/\tau\right),\\
\exp\,({-LSt}) 
&=&I-S+\exp\,(-Lt)S.
\end{eqnarray*}
Substituting them into \eqref{eq_lim1} and using 
\eqref{e_liJ} and Lemma~\ref{l_SJS} in the proof of Theorem\:\ref{th_SpeL} we obtain  
\[
 \lim_{t\to\infty}\exp(-\LT t)
=\lim_{t\to\infty}(S+(I-S)\exp(-t/\tau))(I-S+\vj S)
=S\vj S =\vj S,
\]
\end{remark}
as required.

\begin{remark}
In a number of applications of matrix analysis (see, e.g., \cite{EckartYoung36},\cite[Chapter\:9]{Saaty93}) a matrix is required which is ``the closest'' among the matrices of lower rank 
to a given matrix. The problem discussed in this section can be considered as an inverse one: we construct  matrices $\LT(\tau)$ of rank ${n-1}$ that belong to the class of matrices with zero row sums (which contains the class of Laplacian matrices) and approximate the initial matrix $L$ of lower rank. Consider the following related problem: find $\tau>0$ such that $\LT(\tau)$ is the best mean-squared approximation of~$L$.

Let $\|X\|_E$ be the Euclidean norm of~$X$. Using the symmetry of~$S$ we obtain
\begin{eqnarray}
\!\!\!\!\|\LT(\tau)-L\|^2_E \nonumber
&=&\|\tau^{-1}(I-S)+L S - L\|^{2}_E\\ \nonumber
&=&\trace((\tau^{-1}(I-S)+LS-L)(\tau^{-1}(I-S)+SL^\T-L^\T))\\
&=&\tau^{-2}\trace(I-S)+2 \tau^{-1}\trace((I-S)(LS-L))
+\trace((LS-L)(SL^\T-L^\T)).
\end{eqnarray}

Since for every column $x$ of $L,$ it holds that ${Sx=x}$ (as by Theorem\;\ref{th_domain}, \,$S$ is the projection onto the subspace that contains all columns of~$L$), one has ${SL=L}$ and ${SLS=LS}$, from which it follows that
\[\trace((I-S)(LS-L))=\trace(LS-L-SLS+SL)=0.\]

Using this expression we have:
$$\|\LT(\tau)-L\|^2_E=\tau^{-2}\trace(I-S)+\trace((LS-L)(S L^\T-L^\T)).$$

Since the sum of the eigenvalues of ${I-S}$ is positive (${I-S}$ is a projection all of whose eigenvalues are $1$ and~$0$), we have ${\trace(I-S)>0}$. Consequently, ${\|\LT(\tau)-L\|_E}$ strictly decreases in~$\tau$. On the other hand, the stochasticity of $P$ restricts the growth of $\tau$ by ${(\max_i\sum_{j\ne i}a_{ij})^{-1}}$ (see \eqref{e_P} and~\eqref{e_tau}). Therefore, we have the following result.

\begin{proposition}\label{240514p1} 
${\|\LT(\tau)-L\|_E}$ decreases with the increase of positive $\tau${\rm;} its infimum is ${\|LS-L\|_E}$. Under the condition~\eqref{e_tau} of the stochasticity of~$P,$ the smallest value of ${\|\LT(\tau)-L\|_E}$ is achieved when ${\,\tau=(\max_i\sum_{j\ne i}a_{ij})^{-1}}$.
\end{proposition}

Proposition\,\ref{240514p1} does not imply that $\LT(\tau)=\tau^{-1}(I-S)+LS$ can be equivalently replaced by~$LS$. Indeed, such a replacement does not guarantee consensus, as the rank of $\lim_{t\rightarrow\infty} \exp(-LS t)$ exceeds~$1.$
\end{remark}

\section{CONCLUSION}

In this paper, we considered the orthogonal projection method which provides quasi-consensus when the continuous-time coordination algorithm fails to produce consensus. We gave a characterization of the consensus domain of the basic continuous-time cocoordination algorithm (Theorem\:\ref{th_domain}), proposed an alternative form of the projection method (Section\:\ref{s_ProjA}), and characterized the spectrum of the corresponding matrix $\LT$ (Theorem\:\ref{th_SpeL}). An analogue of the forest consensus theorem for the method of orthogonal projection has been established (Theorem\:\ref{th_ProFo}), a partial result has been obtained on the approximation of the initial coordination algorithm by an algorithm leading to a quasi-consensus (Proposition\:\ref{240514p1}), and an alternative form of the projection method has been obtained for DeGroot's iterative pooling process (Theorem\:\ref{th_AltDe}).

It should be noticed that control problems with dependency digraphs having no spanning in-trees (for example, in the presence of several static or moving leaders \cite{CaoRenEgerstedt12}) have been studied in a number of papers. They were mainly concerned with the problems of satisfying certain space constraints, e.g., entering some localized region of space or retention in it (this line of research is called ``distributed containment control''). However, the problem of reaching consensus without creating additional connections have not been studied in the papers of that trend known to the present authors.

\appendix{}

\vspace*{-1mm}\PTH{\ref{th_domain}}
For the algorithm \eqref{e_modeL} with matrix $L,$ consider protocol \eqref{e_Pm} with matrix $P$ determined by \eqref{e_P} under the assumption that the strict version of \eqref{e_tau} is satisfied. Then by Theorem\:\ref{th_main1}, \eqref{e_AsympStateD2} holds true. From the fact that $\,\lim_{k\to\infty}y(k)$ exists for any $y(0),$ it follows that the powers of $P$ converge. Therefore, by \cite[Theorem~1]{AgaChe11ARCE1} the consensus domain of the algorithm \eqref{e_Pm} is $\TT'(L)=\RR(L)\oplus\spann(\bm1)$. However, according to \eqref{e_AsympStateD2} and \eqref{e_AsympState}, the consensus domains of the algorithms \eqref{e_modeL} and \eqref{e_Pm} coincide, whence $\TT(L)=\RR(L)\oplus\spann(\bm1)$.  \epr

\PPR{\ref{p_STp}}
Since $\TT(L)$ is also the consensus domain $\TT'(L)$ of DeGroot's procedure \eqref{e_Pm}--\eqref{e_P} that satisfies the strict version of
\eqref{e_tau}, we can use the expression for the projection onto $\TT'(L)$ obtained in~\cite[Section\:7]{AgaChe11ARCE1} and coinciding with~\eqref{e_projU}. \epr

\PTH{\ref{th_AltDe}}

The proof is based on the following lemma.

\def\thelemma{A.\arabic{lemma}}

\begin{lemma}\label{le_PS} 
For the matrices $P$ and $S$ determined by Eqs.\ \eqref{e_P} and \eqref{e_projU} and any $k=1,2,\ldots,$ it holds that $(PS)^k=P^kS$.
\end{lemma}

\PLE{\ref{le_PS}}
The trivial ${(PS)^1=P^1S}$ gives the basis of induction (${k=1}$). Let us show that ${SPS=PS}$. Since $S$ is the projection onto $\TT(L)$, for any ${x\in\R^n}$, ${Sx\in\TT(L)}$. Consequently, by Theorem~\ref{th_domain} ${Sx=Ly+\beta\bm1}$ with some ${y\in\R^n}$ and ${\beta\in\R}$. Then ${PSx =(I-\tau L)(Ly+\beta\bm1)=L(I-\tau L)y+\beta\bm1 \in\TT(L)}$ also by Theorem~\ref{th_domain}. Therefore, since $S$ is the projection onto $\TT(L)$, we obtain ${SPSx=PSx}$. Since in the last equality, $x$ is arbitrary, one has ${SPS=PS}$.

The induction step. Assuming that for some ${k\ge1},$ \;${(PS)^k=P^kS}$ and taking into account the identity ${SPS=PS},$ we obtain
${(PS)^{k+1}=(PS)^kPS=P^kS\cdot PS =P^kPS=P^{k+1}S}$.  \epr

Using Lemma\:\ref{le_PS} for the algorithm\:\eqref{e_DeGProjj} with matrix\:\eqref{e_Ptil}, we obtain:
\[x(k)=\P^kx(0)=(PS)^kx(0)=P^kSx(0), \quad k=1,2,\ldots,\]
which coincides with algorithm\:\eqref{e_DeGProj}. \epr

\PTH{\ref{th_SpeL}}
1. Let ${\la\neq0}$ be an eigenvalue of~$L$, let $x$ be a corresponding eigenvector. From ${Lx=\la x}$ it follows that ${x\in\RR(L)}$, and by Theorem\,\ref{th_domain}, ${x\in\TT(L)}$ holds. Since $S$ is the projection onto $\TT(L)$, one has ${Sx=x}$. Therefore, we obtain
\[\LT x=\tau^{-1}(I-S)x+LSx=\tau^{-1}(x-x)+L x=Lx=\la x,\]
i.e., $x$ is an eigenvector of $\LT$ corresponding to the same eigenvalue~$\la.$

Observe that ${\bm1=(1,\ldots,1)^\T}$ is an eigenvector of $\LT$ corresponding to eigenvalue~$0$. Indeed,
\[\LT\bm1=\tau^{-1}(I-S) \bm1+L S \bm1=\tau^{-1}(\bm1-\bm1)+L \bm1=\bf{0}.\]

Due to the diagonalizability of any projection, the eigenvectors of $S$ corresponding to different eigenvalues (these are $0$ and $1$) are linearly independent. Let $y\in\NN(S)$. Since $S$ is symmetric, its rank equals the number of nonzero eigenvalues, i.e., $\rank\xy S=n-d+1$, while the number of zero eigenvalues equals the dimension of~$\NN(S)$. Therefore, it is possible to select $d-1$ of such linearly independent vectors~$y$ and no more. Then, since $Sy=0$, one has
\[\LT y
=\tau^{-1}(I-S)y+LSy
=\tau^{-1}y\]
and 
the multiplicity (both algebraic and geometric) of $\tau^{-1}$ as an eigenvalue of $\LT$ is $d-1$.

2. By item~1, $0$ is a simple eigenvalue of~$\LT$.
\begin{lemma}\label{l_SJS} 
$S\vj S=\vj S$ and $\,\rank(\vj S)=1.$
\end{lemma}

\PLE{\ref{l_SJS}}
For any ${x\in\R^n}$, ${Sx\in\TT(L)}$ by the definition of~$S$. Then by the definition of $\TT(L)$ and the forest consensus theorem, all the components of $\vj Sx$ are the same and, consequently, ${\rank(\vj S)=1}$. Moreover, the algorithm \eqref{e_modeL} does not alter the components of $\vj Sx$ (as it makes the right-hand side of \eqref{e_modeL} vanish), therefore, ${\vj Sx\in\TT(L)}$. Then ${S\vj Sx=\vj Sx}$, since $S$ is a projection onto $\TT(L)$. By the arbitrariness of $x,$ one has ${S\vj S=\vj S}$.
\epr

Using Lemma~\ref{l_SJS} we obtain
\begin{gather*}
\LT(\vj S)
=\tau^{-1}(I-S) \vj S+L S \vj S=\tau^{-1}(\vj S-S \vj S)+L S \vj S
=\tau^{-1}(\vj S- \vj S)+L \vj S
={\bf 0}
\end{gather*}
and
\[(\vj S)\LT=\tau^{-1}(\vj S-\vj S S) +\vj S L S =\vj S L S={\bf 0}.\]
According to Theorem~\ref{th_SpeL}, $\rank\,\LT=n-1$, therefore, $\ind\LT=1$. Since $\rank(\vj S)=1$, we conclude (see \cite{Wei96} or the characterization (b) in \cite{AgaChe02}) that $\vj S$ is the eigenprojection of~$\LT$.

3. Having in mind that the orthogonal projection $S$ is symmetric, consider the matrix 
$$\LT\LT^\T=(\tau^{-1}(I-S)+L S)(\tau^{-1}(I-S)+S L^\T)=\tau^{-2}(I-S)+L S L^\T.$$ 
Suppose that $\rank S<n$ and $\,y\in\NN(S)$. Since $S$ is an orthogonal projection, the elements of the null space of $S$ are orthogonal to the range of $S$, which by Theorem~\ref{th_domain} includes all the columns of~$L$: $L^\T y=\bm0$. Then
\[
\LT\LT^\T y
=\tau^{-2}(I-S)y+L S L^\T y
=\tau^{-2}y,
\]
i.e., $ \tau^{-1}$ is a singular value of~$\LT$. Its multiplicity is $d-1$, since it is the dimension of $\NN(S)$, from which the vectors~$y$ are drawn (see the proof of item~1 of this theorem).  \epr

\PTH{\ref{th_ProFo}}
By item~1 of Theorem~\ref{th_SpeL}, $0$ is a simple eigenvalue of~$\LT$, while the real parts of all other eigenvalues, as well as for $L$, are positive. The rest of the proof is carried out (using item~2 of Theorem~\ref{th_SpeL}) analogously to the proof of Theorem~\ref{th_main}. \epr

%% file: CALL.bbl
\begin{thebibliography}{24}
\itemsep0mm
\parsep0mm

\bibitem{AgaChe00}
  {Agaev,~R.P. and Chebotarev,~P.Yu.},
  {The Matrix of Maximum {Out} Forests of a Digraph and its Applications}, 
  \emph{Automat. Remote Control.},
  2000,
  vol.~61, no.~9,
  pp.~1424--1450.

\bibitem{CheAga09ARC}
  {Chebotarev,~P.Yu. and Agaev,~R.P.},
  {Coordination in Multiagent Systems and {Laplacian} Spectra of Digraphs}, 
  \emph{Automat. Remote Control.},
  2009,
  vol.~70, no.~3,
  pp.~469--483.

\bibitem{Olfati-SaberMurray04}   
  {Olfati-Saber,~R. and Murray,~R.M.},
  {Consensus Problems in Networks of Agents with Switching Topology and Time-delays}, 
  \emph{IEEE Trans. Automat. Control.},
  2004,
  vol.~49, no.~9,
  pp.~1520--1533.

\bibitem{MesbahiEgerstedt10book}   
  {Mesbahi,~M. and Egerstedt,~M.},
  \emph{Graph Theoretic Methods in Multiagent Networks}.
  Princeton: Princeton University Press, 2010. 

\bibitem{CheAga13UBSr}  
  {Chebotarev,~P.Yu. and Agaev,~R.P.},
  {On the Asymptotics of Consensus Protocols}, 
  \emph{Large-scale Systems Control},
  2013,
  vol.~43,
  pp.~55--77 (in Russian).

\bibitem{CheAga14IEEETAC}   
  {Chebotarev,~P. and Agaev,~R.},
  {The Forest Consensus Theorem},
  \emph{IEEE Trans. Automat. Control.},
  2014,
  vol.~59, no.~9,
  pp.~2475--2479.

\bibitem{AgaChe11ARCE1}
  {Agaev,~R.P. and Chebotarev,~P.Yu.},
  {The Projection Method for Reaching Consensus and the Regularized Power Limit of a Stochastic Matrix},
  \emph{Automat. Remote Control.},
  2011,
  vol.~72, no.~12,
  pp.~2458--2476.

\bibitem{CheAga02ap}   
  {Chebotarev,~P. and Agaev,~R.},
  Forest Matrices around the {Laplacian} Matrix, 
  \emph{Linear Algebra and its Applications.},
  2002,
  vol.~356,
  pp.~253--274.

\bibitem{DeGroot74}   
  {DeGroot,~M.H.},
  {Reaching a Consensus}, 
  \emph{J. Amer. Statistical Association},
  1974,
  vol.~69, no.~345,
  pp.~118--121.

\bibitem{Gantmacher60}
  {Gantmacher, F.R.},
  \emph{The Theory of Matrices}.\hskip 1em plus 0.5em minus
  0.4em\relax New York: Chelsea, 1959.

\bibitem{EckartYoung36}   
  {Eckart,~C. and Young,~G.},
  {The Approximation of One Matrix by Another of Lower Rank}, 
  \emph{Psychometrika},
  1936,
  vol.~1, no.~3,
  pp.~211--218.

\bibitem{Saaty93} 
  {Saaty,~T.L.},
  \emph{The Analytic Hierarchy Process}.
  Pittsburgh: RWS Publications, 1980.

\bibitem{CaoRenEgerstedt12}   
  {Cao,~Y., Ren,~W. and Egerstedt~M.},
  {Distributed Containment Control with Multiple Stationary or Dynamic Leaders in Fixed and Switching Directed Networks}, 
  \emph{Automatica},
  2012,
  vol.~48, no.~8,
  pp.~1586--1597.

\bibitem{Wei96}   
  {Wei,~Y.}, 
  {A Characterization and Representation of the {Drazin} Inverse}, 
  \emph{SIAM J. Matrix Analysis Applications},
  1996,
  vol.~17, no.~4,
  pp.~744--747.

\bibitem{AgaChe02}
  {Agaev,~R.P. and Chebotarev,~P.Yu.},
  {On Determining the Eigenprojection and Components of a Matrix}, 
  \emph{Automat. Remote Control},
  2002,
  vol.~63, no.~10,
  pp.~1537--1545.

\end{thebibliography}
